\documentclass{article}
\usepackage{graphicx}
\usepackage{subcaption}

\usepackage{natbib}
\usepackage{amsfonts}
\usepackage{amssymb}
\usepackage{amsthm}
\usepackage{latexsym}
\usepackage{xspace}

\usepackage{amsmath}
\usepackage[latin1]{inputenc}

\usepackage{mathrsfs}
\usepackage{amsmath}

\newcommand{\bmu}{\boldsymbol{\mu}}

\newcommand{\bsigma}{\boldsymbol{\sigma}}

\newcommand{\bone}{\boldsymbol{1}}

\newcommand{\ba}{\boldsymbol{a}}
\newcommand{\bb}{\boldsymbol{b}}
\newcommand{\bm}{\boldsymbol{m}}

\newcommand{\bc}{\boldsymbol{c}}
\newcommand{\bu}{\boldsymbol{u}}

\newcommand{\bp}{\boldsymbol{p}}

\newcommand{\bpsi}{\boldsymbol{\psi}}

\newcommand{\bzero}{\boldsymbol{0}}

\begin{document}

\title{Efficient Breeding by Genomic Mating}

\author{Deniz Akdemir  \footnote{Corresponding author: deniz.akdemir.work@gmail.com}\\ Department of Plant Breeding \& Genetics\\ 
  Cornell University\\ Ithaca, NY USA \\
Julio Isidro S\'anchez \\ Agriculture \& Food Science\\ 
 University College Dublin\\ Dublin, Ireland}

\maketitle

\begin{abstract}
In this article, we propose an approach to breeding which focuses on mating instead of truncation selection, our method uses  genome-wide marker information in a similar fashion to genomic selection so we refer it to as genomic mating. Using concepts of estimated breeding values, risk (usefulness) and inbreeding, an efficient mating approach is formulated for improvement of breeding values in the long run. We have used a genetic algorithm to find solutions to this optimization problem. Results from our simulations point to the efficiency of genomic mating  for breeding complex traits compared to truncation selection.

\end{abstract}


{Keywords \& Phrases: Breeding, phenotypic selection, genomic selection, genomic mating, complex traits, genome-wide markers, inbreeding, genomic diversity, portfolio optimization}

Selection is an evolutionary phenomenon that affects the phenotypic distribution of a population. From a breeding point of view, truncation selection means breeding from the ''best'' individuals \citep{falconer1996introduction}. Breeders have been selecting on the basis of phenotypic values since domestication of plants and animals or, more recently,  breeders have substantially used the pedigree-based prediction of genetic values for the genetic improvement of complex trait \citep{henderson1984applications, gianola1986bayesian, crossa2006modeling, piepho2008blup};  this is called phenotypic selection (PS).

Since the invention of the polymerase chain reaction by  Mullis in 1983, the enhancements in high throughput genotyping \citep{lander2001initial,margulies2005genome,metzker2010sequencing} have transformed breeding pipelines through marker-assisted selection (MAS) \citep{lande1990efficiency}, marker assisted introgression \citep{charcosset1997marker}, marker assisted recurrent selection \citep{bernardo2006usefulness}, and genomic selection (GS) \citep{Meuwissen1819}. The latter use genome-wide markers to estimate the effects of all genes or chromosome positions simultaneously \citep{Meuwissen1819} to calculate genomic estimated breeding values (GEBVs), which are used for selection of individuals. This process involves the use of phenotypic and genotypic data to build prediction models that would be used to estimate GEBV's from genome wide marker data. It has been proposed that GS increases the genetic gains by reducing the generation intervals and also by increasing the accuracy of estimated breeding values. However, many factors are involved in the relative per unit of time efficiency of GS and its short and long time performance \citep{jannink2010genomic,daetwyler2007inbreeding}.

Some optimized parental contribution calculation schemes have been proposed to balance the gain from selection and variability \citep{wray1994moet,  brisbane1995balancing,  meuwissen1997maximizing, Meuwissen1819,sonesson2012genomic, clark2013effect}. Approaches that  seek for an optimal subset of mates among potential male and female candidates have been formulated from an animal breeding perspective in \cite{allaire1980mate, jansen1985selecting, kinghorn1998mate} and in subsequent articles \citep{kinghorn1999mate, fernandez2001practical, berg2006eva, kinghorn2011algorithm, pryce2012novel, sun2013mating}. These approaches also seek solutions that attain a balance between genetic gains and inbreeding and most developments in this area have been focusing on animals. 
 
\begin{figure*}
	\centering
		\includegraphics[width=1\textwidth]{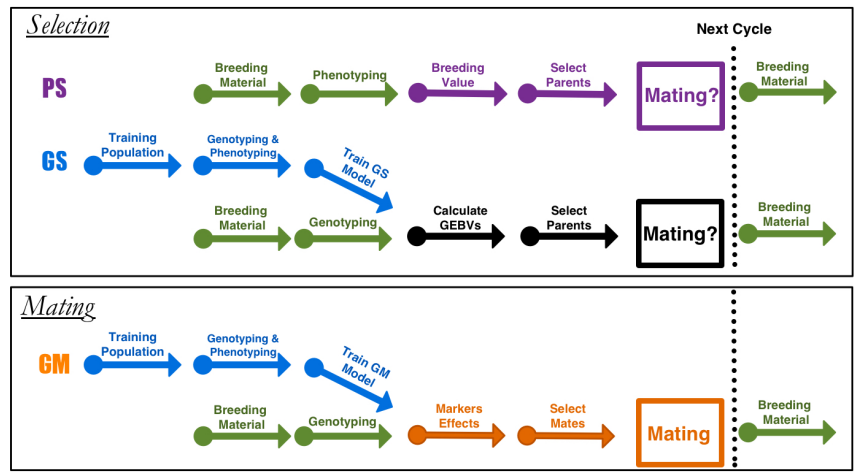}
	\caption{Diagram for the different breeding approaches. Phenotypic selection (PS) and genomic selection (GS) are truncation selection methods, and genomic mating (GM) is the mating approach. Arrows indicate the different stages in a breeding cycle. In PS, starting with a set of parents as breeding material, selection is performed based on phenotypes. In GS, the breeding value is predicted using a statistical model based on phenotypes and whole genome marker data (obtained within an experiment that is repeated in every few cycles, blue arrows) selection is based on GEBVs. Genomic mating is similar to GS in terms of estimating marker effects, but in GM the genetic information and the estimated marker effects are used to decide for the list of mates that should be crossed to obtain the next breeding population. Genomic mating is the only approach that gives an answer to the mating question: ''Who is mating whom?''.}
	\label{fig:fig1}
\end{figure*}

Marker assisted breeding to stack genes using complementary crosses has been useful for breeders when the trait of interest is regulated by only a few loci. For complex traits, on the other hand, there is a scarcity of methods available to breeders.  Both of PS and GS focus on improvement by truncation selection, mainly ignoring the role of mating and complementation as an evolutionary force (Figure \ref{fig:fig1}). For this reason both PS and GS are, in a sense, inefficient for improving complex traits in the long term.  Methods that seek only a balance between genetic gains and inbreeding are incomplete because they ignore the variances in the genetic values; measures of gain do not completely capture the full potential of a mate pair.

In this article, we propose an optimal genomic mating (GM) approach for breeding (Figure \ref{fig:fig1}). Our approach is focused on mainly on plant breeding scenarios. We believe it uses genomic information more completely than the recently proposed genomic selection and reinforces mating complementary individuals. Given a set of individuals in the current breeding population, their corresponding markers and related marker effects, our solution is a list of mates that should be crossed for obtaining the next breeding population instead of a list of individuals in the current breeding population which will become the parents of the next generation. Unlike selection methods, GM approach does not exclude the possibility of contribution of all individuals to the next generation. A cross-variance term is included in the objective function along with genetic gains and inbreeding to account for potential benefits from including mates with higher estimated genetic variance. To this end, we provide a method that uses marker effect estimates to estimate within cross-variances assuming independence among loci and additive effects. The difficult computational problem of finding the optimal set of mates have been handled by an efficient genetic algorithm. Using simulations, we have compared the long range performance of GM to PS, GS and an optimal parentage contribution approach.  Results from these simulations point to the viability and efficiency of genomic mating  for breeding complex traits.

\section*{Methods}

It is widely accepted that short term gains from selection increases with increased selection intensity. However, increasing selection reduces the genetic variability, which increases the rates of inbreeding and may reduce gains in the long term run. Most of the selection in plant breeding are designed to maximize genetic gain but some approaches try to balance the gain from selection and variability. We will give a brief review of these approaches since they relate to the mating theory. 

\subsection*{Current methodology.}

Many authors \citep{goddard2009genomic, jannink2010dynamics,sonesson2012genomic, sun2013mating, clark2013effect} have expressed the importance of reducing inbreeding in PS and GS for long-term success. They argued that GS is likely to lead to a more rapid decline in the selection response unless new alleles are continuously added to the calculation of GEBVs, stressing the importance of balancing short and long term gains by controlling inbreeding in selection.

Let $A$ be a matrix of pedigree based numerator relationships or the additive genetic relationships between the individuals in the genetic pool (this matrix can be obtained from a pedigree of genome-wide markers for the individuals) and let $\bc$ be the vector of proportional contributions of individuals to the next generation under a random mating scheme. The average relatedness for a given choice of  $\bc$ can be defined as $r=\frac{1}{2}\bc'A\bc.$ If $\bb$ is the vector of GEBV's, i.e., the vector of BLUP estimated breeding values of the candidates for selection. The expected gain is defined as $g=\bc'\bb.$ Without loss of generality, we will assume that the breeders long term goal is to increase the value of $g.$ 

In \citep{wray1994moet,  brisbane1995balancing,  meuwissen1997maximizing} an approach that seeks maximizing the genetic gain while restricting the average relationship is proposed. The optimization problem can be stated as

\begin{equation}\begin{array}{lc}
\underset{\bc}{\text{minimize}}& r=\bc' \frac{A}{2} \bc \\ [10pt]
\mbox{subject to} & \begin{array}[t]{rcl}
\bc' \bb=\rho  &  \\ [10pt]
\bc'\bone=1 & \\ [10pt]
\bc\geq 0 &
\end{array}\end{array}\label{eq:1}\end{equation}

This problem is easily recognized as a Quadratic Optimization problem (QP).  There are many efficient algorithms that solves QP's so there is in practice little difficulty in calculating the optimal solution for any particular data set. Recently, several allocation strategies were tested using QP's in \citep{goddard2009genomic, pryce2012novel, schierenbeck2011controlling}. It is easy to extend these formulations to introduce additional constraints as positiveness, minimum-maximum for proportions, minimum-maximum for number of lines (cardinality constraints).

Some authors recommended mate selection approaches that also seek a balance between gain and inbreeding from an animal breeding perspective \citep{allaire1980mate, jansen1985selecting, kinghorn1998mate}. Kinghorn in a series of articles \citep{kinghorn1998mate,kinghorn1999mate, kinghorn2011algorithm} describes an algorithmic approach that separates the optimization and the objective function for the mate selection approach and therefore can be used for a wide array of optimization criteria (mate selection index) with hard and soft constraints. Similar algorithmic approaches were recommended in \cite{fernandez2001practical, pryce2012novel,sun2013mating}.  However, none of these methods include a term for the genotypic variance of the crosses, such as described in this paper. 

By solving the QP in (\ref{eq:1}) for varying values of $\rho,$ or using the similar criteria in the mate selection approaches, we can trace out an efficient frontier curve, a smooth non-decreasing curve that gives the best possible trade-off of genetic variance against gain. This curve represents the set of optimal allocations and it is called the efficiency frontier (EF) curve in finance \citep{markowitz1952portfolio} and breeding literature.

 \subsection*{Optimal genomic mating.}

There are several alternative measures of inbreeding based on mating plans \citep{leutenegger2003estimation, wang2011coancestry}. In this article, we have used a measure derived under the  infinitesimal genetic effects assumption proposed by \citep{quaas1988additive} and \citep{legarra2009relationship}. A measure of gain, i.e., the total expected breeding value of the progeny, can also be calculated from the results of the same authors. However, in our belief, the expected value by itself is not a good measure of possible gains since it carries no information about the variability of breeding values (BV's) among full-sibs.  Therefore, we have derived a measure called the risk  of a mating plan (this is related to the concept of ''usefulness'')  by increasing the expected BV's of the progenies by a small amount (the intensity is controlled by the parameter $\lambda_1$) proportional to their expected variance (standard deviation) calculated under the  infinitesimal effects assumption. Other measures of expected variance could also been used. For example, it is possible to calculate this variance by simulating progenies for parent pairs, and one can easily include information about the LD in these simulations. Another measure of risk was proposed in \citep{zhong2007using}. The measures of inbreeding and risk  we chose are computationally efficient and this makes the optimization over the mates feasible. 

Combining the measures of inbreeding and risk into one leads to the formulation of the mating problem:
\begin{equation}\label{eq:eqoptprob}
\underset{P_{32}}{\text{minimize}} \quad  r( \lambda_1, \lambda_2, P_{32}) =  -Risk(\lambda_1, P_{32}) + \lambda_2 * Inbreeding(P_{32})
\end{equation}
where $\lambda_2\geq 0$ is the parameter whose magnitude controls the amount of inbreeding in the progeny, and the minimization is over the space of the mating matrices $P_{32}.$  $\lambda_1$ controls allele heterozygosity weighted by the marker effects and  $\lambda_2$ controls allele diversity. When $\lambda_1=0$ the risk measure is the same as total expected gain. 

Now, we give the details of how the measures $Risk(\lambda_1)$ and $Inbreeding$ are defined in this paper. Let $\bb=(\bb'_1, \bb'_2, \bb'_3)'$ denote the vector of genetic effects corresponding to the parents and progeny, where $\bb_1$ and $\bb_2$ are the genetic effects of the $N$ parents and $\bb_3$ are the genetic effects of the $N_c$ progeny. Let the pedigree based numerator relationship matrix for the individuals in $\bb$ be $A$ and $A$ is partitioned  as \[ A = \begin{bmatrix} A_{11}&A_{12} & A_{13}\\ A_{21}&A_{22} & A_{23} \\ A_{31} & A_{32} & A_{33} \end{bmatrix}\] corresponding to the partitions of $\bb.$ Suppose, we also have the markers for the parents in the second partition, and $\bu_2=M \ba$ where $M$ is the matrix of minor allele frequencies, coded as $0,1,$ and $2.$ Let $M_c$ be the $N\times m$ marker allele frequency centered incidence matrix ($M_c=M-2\bone_N(p_1, p_2,\ldots,p_m)$) and $\ba$ be the vector of marker effects. Variance-covariance of $\bb_2$ can be written as \[Var(\bb_2)= \frac{M_cM_c}{k}\sigma^2_b=G\sigma^2_b\] where $k=\sum_{j=1}^{m}2p_j(1-p_j)$ is twice the sum of heterozygosities  of the markers \citep{vanraden2008efficient}.

Following \cite{quaas1988additive} and \cite{legarra2009relationship}, let $P$ be a matrix containing the transitions from ancestors to offspring. We will refer $P$ as the mating or parentage matrix. Then, we can write $\bb=P\bb+\bpsi$ where  $\bpsi$ is the vector of Mendellian samplings and founder effects with a diagonal variance D. In particular, using only the rows of $P$ corresponding to the $\bb_3$ the relationship is written as \[\bb_3=\begin{bmatrix} P_{31} & P_{32} & P_{33}\end{bmatrix}\begin{bmatrix}\bb_1 \\ \bb_2 \\ \bb_3 \end{bmatrix}+\bpsi_3\] which can also be stated as a regression equation of the form $ \bb_3 =(I-P_{33})^{-1}(P_{31}\bb_1 + P_{32}\bb_2+\bpsi_3)$ \citep{quaas1988additive}. The variance-covariance matrix of $\bb_3$ is given by \begin{equation}\label{covofb3} Var(\bb_3)=(I-P_{33})^{-1} (P_{31}A_{11}P'_{31}+P_{32}G P'_{32}+P_{32}A_{21}P'_{31}+P_{31}A_{12}P'_{32}+D_3) (I-P_{33})'^{-1} .\end{equation}

The variances caused by Mendelian sampling in $D_3$ are related to inbreeding in the parents via \[var(\psi)\propto (1/2-(F_1+F_2)/4)\] where $F_1$ and $F_2$ are the inbreeding coefficients of the two parents which can be extracted from the diagonals of G. The variance-covariance formula reduces to  \[Var(\bb_3)=P_{32}G P'_{32}+D_3\] if all the founders are genotyped (no $P_{31}$), and a relatively simple mating strategy is assumed where founders are the only parents and no back-crossing is allowed ($P_{33}=\bzero$).  This is the assumption made for the remainder of this paper and in this case $ P_{32}$ is a $N_c \times N$ matrix ($N_c$ children from $N$ parents) with each row having two $1/2$ values at positions corresponding to two distinct parents or only a value of 1 at the position corresponding to the selfed  parent. All the other elements of this matrix are zero. Nevertheless, one can easily imagine situations where some of the founders are not genotyped or when some of the progeny also have progeny, then the formula in (\ref{covofb3}) will be relevant. If some founders are not genotyped but a pedigree is available relating them to the rest of the founders then  the variance-covariance for the founders, $Var(\bb_1,\bb_2),$ can be calculated using the relationship matrix in \cite{legarra2009relationship}. Furthermore, construction of the mating matrices for more complex mating plans is described in \citep{quaas1988additive}.

$Var(\bb_3)$ gives us the expected variance-covariance of the progeny given the mating matrix $ P_{32}$ and the realized relationship matrix $G$ of the parents. This can be used as to measure the expected genetic diversity of a mating plan: We can use a measure in line with the inbreeding term $\bc' A \bc$ in (\ref{eq:1}) by \[Inbreeding(P_{32})= \bone_{N_c}'Var(\bb_3)\bone_{N_c}=\bone_{N_c}'(P_{32}G P'_{32}+D_3)\bone_{N_c}.\]

We also need a measure for genetic gain. A simple measure of gain for a given mating plan expressed in $P_{32}$ can be constructed from the expected value of $\bb_3:$ \[E(\bb_3)=P_{32}M\ba\] and an overall measure can be written as \[Gain(P_{32})=\bone_{N_c}'E(\bb_3).\] 

Finally, we want to complement the measure ''gain'' with a measure of within cross-variance for the genetic levels of children of the parent pairs. Suppose the organism under study is diploid. We can recode the markers matrix $M$ coded as -1, 0, and 1 into a $N\times m$ matrix $M^*$ using the information in the marker effects vector $\ba$ such that markers are coded as the number of beneficial alleles as 0,1, or 2. This is achieved by first obtaining marker effects estimates and then using the sign of the estimates to determine what is a beneficial allele. We can also obtain a related marker effects vector $\ba^*$ by replacing the original marker effects by the effects of the beneficial alleles ($\ba^*=|\ba|$) so that we have $M\ba=(M^*-\bone_{N\times m})\ba^*.$ For a given parent pair, we can calculate the vector expected number of beneficial alleles of the children of these parents using a transition vector $\bp$ as $\bmu=E(\bm)=\bp'M^*.$ In addition, for each locus we can calculate the variance for the number of beneficial alleles from the number of alleles the parents have and put them in a vector which we will denote by $\bsigma_{\bp}=(\sigma_{\bp 1},\sigma_{\bp 2},\ldots, \sigma_{\bp m}).$ Calculation of elements of  $\bsigma_{\bp}$ from the coding in $M^*$ can be as in Table \ref{tab:tab1}.  We define  risk measure for this parent pair as \[Risk(\lambda_1) = (\bp' M^* +\lambda_1*\begin{pmatrix}\sqrt{\sigma_{\bp 1}}\\ \sqrt{\sigma_{\bp 2}} \\ ... \\ \sqrt{\sigma_{\bp m}} \end{pmatrix}-\bone_m)'\ba^*\] where $\lambda_1\geq 0$ is the risk parameter and $m$ is the number of markers. The risk of a mating plan (which is expressed in $P_{32}$) is the sum of all the risk scores for all mate pairs in that plan which we will denote by $Risk(P_{32},\lambda_1).$

\begin{table}
\centering
\caption{Calculation of mean number and variance of the beneficial alleles of progeny at each locus from the beneficial allele code (-1, 0, 1) of the parents at the same locus.}
\label{tab:tab1}
\begin{tabular}{|l|l|l|l|}
\hline
\textbf{Parent 1} & \textbf{Parent 2} & \textbf{Expected Number of Beneficial Alleles} & \textbf{Variance} \\ \hline
1                 & 1                 & 2                             & 0                 \\ \hline
1                 & 0                 & 1.5                           & 0.25               \\ \hline
0                 & 1                 & 1.5                           & 0.25               \\ \hline
1                 & -1                & 1                             & 0                 \\ \hline
-1                & 1                 & 1                             & 0                 \\ \hline
0                 & 0                 & 1                             &0.5               \\ \hline
0                 & -1                & 0.5                           & 0.25               \\ \hline
-1                & 0                 & 0.5                           & 0.25               \\ \hline
-1                & -1                & 0                             & 0                 \\ \hline
\end{tabular}
\end{table}

If the risk parameter $\lambda_1$ is set to zero then we have \[Risk(P_{32},\lambda_1=0)=\bone_{N_c}'E(\bb_3)=\bone_{N_c}'P_{32}M\ba.\] The magnitude of $\lambda_1$ is related to the desire of the breeder to take advantage of within cross variances and encourages mating parents that are heterozygotes at QTL.

In this sense, the efficient mating problem can be stated as an optimization problem as follows: 
\begin{equation}\begin{array}{lc}
\underset{P_{32}}{\text{minimize}}&Inbreeding(P_{32})= \bone_{N_c}'(P_{32}G P'_{32}+D_3)\bone_{N_c}\\ [10pt]
\mbox{subject to} & \begin{array}[t]{rcl}
Risk(P_{32},\lambda_1)=\rho.  &
\end{array}\end{array}\label{eq:4}\end{equation} In the above optimization problem, we are trying to minimize the inbreeding in the progeny while the risk is set at the level $\rho\geq 0.$ In the remainder of this paper, we will  use the the following equivalent formulation of the mating problem in Equation (\ref{eq:eqoptprob}).

The optimization problem in (\ref{eq:eqoptprob}) is a combinatorial problem whose order increases with the number of individuals in the breeding population and the number of progeny. We have devised a genetic algorithm to tackle this optimization problem and found that the algorithm is very efficient for finding good solutions in reasonable computing time. Genetic algorithms (\cite{ holland1973genetic, davis1991handbook, goldberg2006genetic}) are particularly suitable for optimization of combinatorial problems.  The idea is to use a population of candidate solutions that is evolved toward better solutions. At each iteration of the algorithm, a fitness function is used to evaluate and select the elite individuals and  subsequently the next population is formed from the elites by genetically motivated operations like crossover and mutation. It should be noted that the solutions obtained by a genetic algorithm will usually be sub-optimal and different solutions can be obtained given a different starting population of candidate solutions.  We did not explore any alternatives to our mating optimization algorithm, but similar evolutionary algorithms like differential evolution, particle swarm, tabu search, and simulated annealing or hill climbing methods like the exchange method can be useful to solve this problem.  As stated by other authors  \cite{kinghorn2011algorithm} and \cite{pryce2012novel}, the mate selection problem has two independent components: A mate selection index (MSI), i.e., the optimization function and a mate selection algorithm that can be used to optimize the MSI. In our article, we have provided new approaches to both of these components: First, the MSI we have used differed from previous authors and included terms for gain, variance  and inbreeding, and secondly, we have adopted a genetic algorithm that can efficiently look for good solutions. 

As opposed to the continuous parentage contribution proportions solutions in the GS method, the mating method gives discrete solutions. That is to say, the solutions of the mating algorithm are the list of parent mates of the progeny. Additionally,  there is no real guideline for choosing where to operate while using GS method. Conversely, since the mating algorithm is discrete and the number of genotypes contributing to the next generation increase starting from one as we increase the $\lambda_2$, we can identify a point to operate on this surface by slowly increasing the   $\lambda_2$ until a desired minimum number of  genotypes are included in the solution. This is the method we have used in our simulations where we have run several cycles of mating. We included the minimum number of parents as a parameter: ''minparents'' in  simulations. This allowed us to run the simulations many times without interference. However, a better approach in practical situations would be to plot the whole frontier surface and select a solution that has a good risk to diversity ratio.

There is an intrinsic limit to the amount of selfing or crosses of closely related lines in GM.  Although it is hard to imagine that this is what is done in practice, theoretically, leaving the decision to a ''roulette wheel'' assignment of parents as mates as in the selection approach might lead to too much inbreeding. For example, if the parental contribution proportion of a parent is $0.50,$ then we expect to have $25\%$ obtained by selfing this parent. GM allows a better control of inbreeding by completely controlling who mates with whom. 

\section*{Results}

\begin{figure*}	
\centering
      \includegraphics[width=1\textwidth,angle=0]{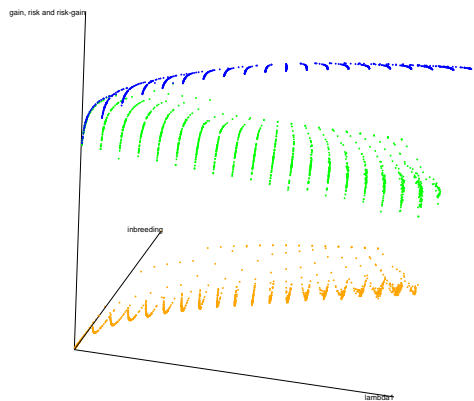}

 \caption{A marker data was created for 50 genotypes by randomly generating 1000 markers for each genotype. By introducing independent and identically normally distributed marker effects at 500 of randomly selected the loci we have defined a trait.  Three surfaces are given in the figure. The blue surface represents the optimal values of the objective function in  Equation (\ref{eq:eqoptprob}) Points below this surface correspond to sub-optimal regions and points above this surface are unattainable. The points along the surfaces are the optimal points balancing gain, risk and inbreeding. The green surface is the expected average genetic value of the progeny and the orange surface is the value of the cross-variance term, these two surfaces add up to the blue surface.}
 \label{fig:fig2}
\end{figure*}

For a set of 50 simulated lines, we have identified optimal mates for the progeny at changing values of $\lambda_1$ and $\lambda_2.$ The frontier surface is drawn using the optimal mating algorithm (Figure \ref{fig:fig2}). The coordinates of the  points on the curve are the values of estimated risk, inbreeding  and the difference between risk and gain. for the optimal sets of mates. The blue surface represents the optimal values of the objective function in  Equation (\ref{eq:eqoptprob}) Points below this surface correspond to sub-optimal regions and points above this surface are unattainable. The points along the surfaces are the optimal points balancing gain, risk and inbreeding. The green surface is the expected average genetic value of the progeny and the orange surface is the value of the cross-variance term, these two surfaces add up to the blue surface. By changing $\lambda_1$ and $\lambda_2$ we move on this surface. Since the points on this surface correspond the optimal solutions, the breeder should operate on the surface. The optimal solutions to the mating problem at a few selected values of $\lambda_1$ and $\lambda_2$ are in Figures 3a-3d.

Efficient frontier surface is the basis for GM. A feasible mating plan is one that meets specified constraints. The EF surface allows breeders to understand how a mating plan's expected risk vary with the amount of inbreeding. However, the decision depends on how much more or less risk a breeder wants to take. Most breeders will be willing to assume a greater inbreeding for a greater risk. Breeders differ in the amount of inbreeding they are willing to take for a given risk. Breeders who are inbreeding averse require lower inbreeding for a given amount of risk than breeders who are risk seekers. 

\begin{figure*}
\centering 
     \includegraphics[width=.9\textwidth]{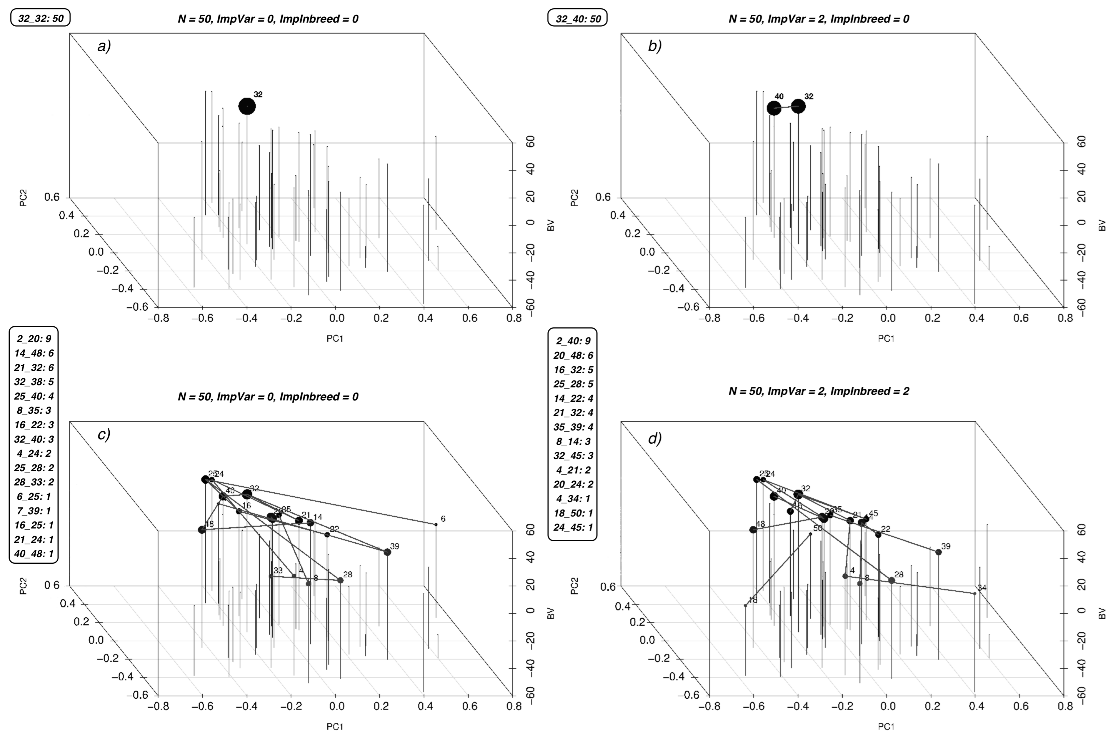}
 \label{fig:fig2.2}
 \caption{Optimal solutions to the mating problem at a few selected values of $\lambda_1$ and $\lambda_2$ are in (a), (b), (c), and (d). The list of mates and the number of crosses for each mate is given along the figures. The first two coordinates are used to display the genetic relationships of the lines using the first two principal components, the third coordinate displays the breeding values of the parents.  Each parent is represented by a vertical bar. The lines between the vertical bars represent the matings and the size of the points on the bars are proportional to the number of crosses between that parent and any other.}
\end{figure*}

\begin{figure*}
\centering
\begin{subfigure}{.5\textwidth}
  \centering
  \includegraphics[width=.70\linewidth, angle=270]{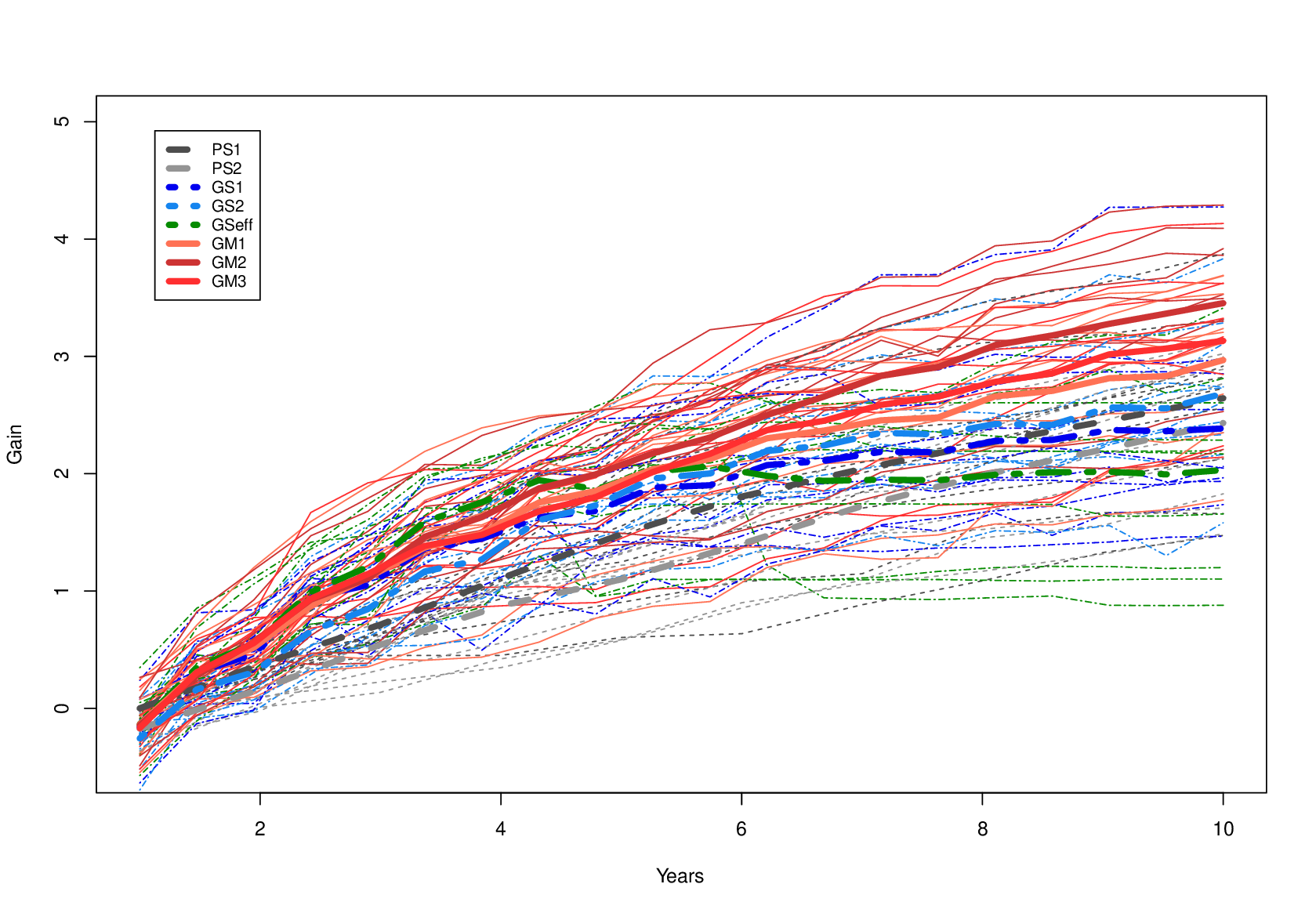}
  \caption{150 lines}
  \label{fig:fig3}
\end{subfigure}%
\begin{subfigure}{.5\textwidth}
  \centering
  \includegraphics[width=.70\linewidth, angle=270]{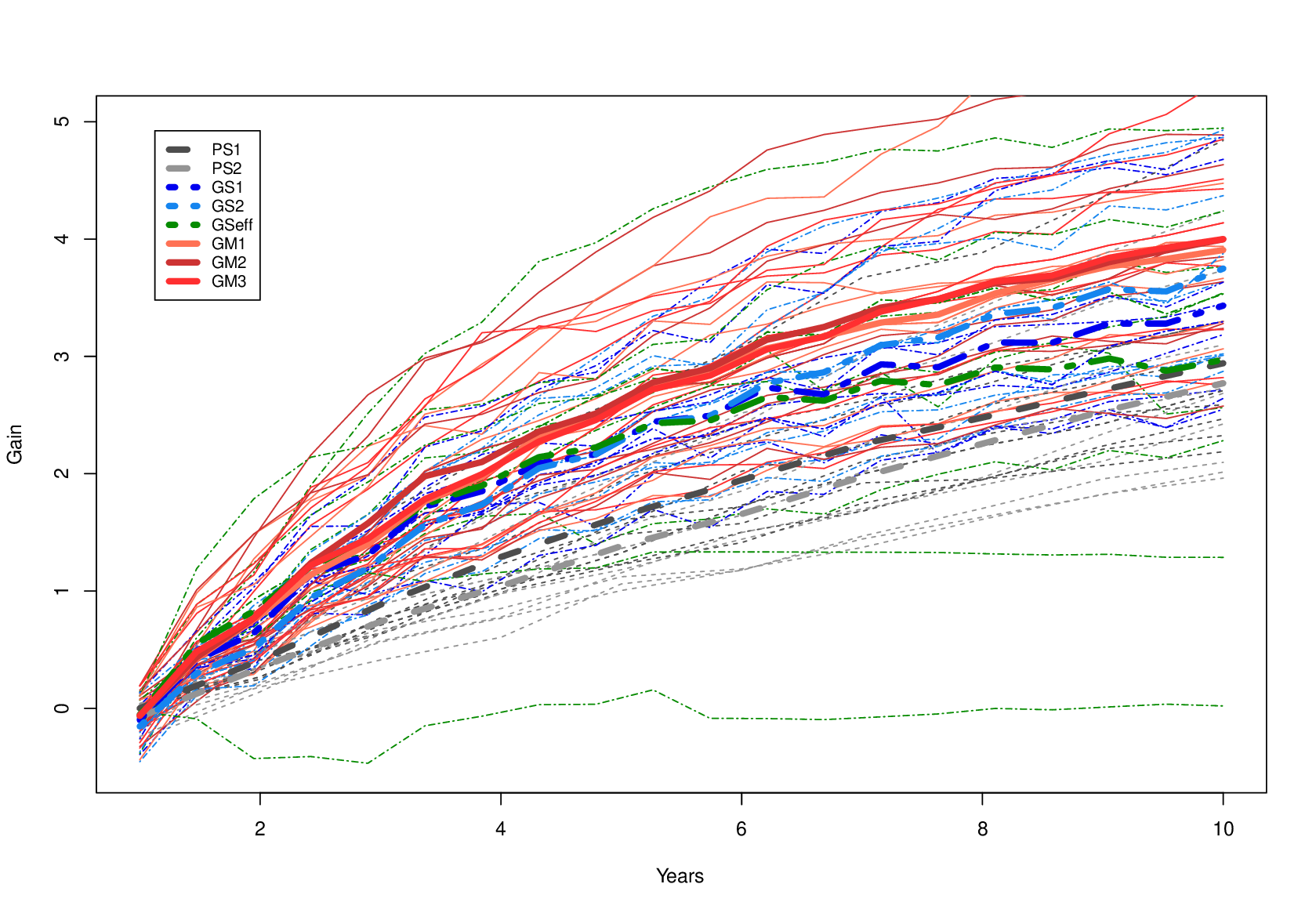}
  \caption{300 lines}
\label{fig:fig4}
\end{subfigure}
\caption{The long term behavior of PS, GS, Efficient GS and GM. Starting from 2 founders we have formed a population of 150 (\ref{fig:fig3}) and 300 (\ref{fig:fig4}) genotypes with 1000 SNPs at 3 chromosomes each and carried this population through 200 generations of random mating and 100 generations of phenotypic selection based on a complex trait (300 QTL at random locations on each chromosome) with 0.5 heritability generated based on the infinitesimal model. Starting from this initial population, we have simulated 10 rounds of PS, and 20 rounds of GS and GM (assuming one cycle of PS and two cycles of GS and GM per year). For GS and GM, the marker effects were estimated from data once per year. The results of 10 replication of this simulation with selection intensity $10\%$ (PS1, GS1) and $20\%$ (PS2, GS2) for PS and GS; Efficient GS (GSeff); and GM with $\lambda_2=0, 5, 10$ (GM1, GM2, GM3).  Each thin line represents the genetic gains over cycles by different methods over a replication of the experiment. The thick lines show the mean improvement for each of the methods over 10 replications. In these simulation studies there is a clear advantage of using GM as a breeding method.}
\label{fig:test}
\end{figure*}

Figure \ref{fig:fig3} and \ref{fig:fig4} show the results from  simulations for the study of the long term behavior of PS, GS, and GM.  In this simulation study, there is a clear advantage of using GM as a breeding method.

\section*{Discussions}
In this article, we have proposed a new methodology for breeding living organisms based on optimal genomic determination of mating plans. Our approach can be contrasted with the selection approach where only proportional contributions of parents to the progeny are the main focus. A major novelty in GM approach as compared to the other methods is the utilization of within cross-variances (usefulness) in the objective function along with genetic gains and inbreeding. 

Although similar to GS in its information requirements, our approach offers a better utilization of the genotypic and phenotypic information. Under the optimal mating breeding scheme some concepts in statistical genetics like selection intensity needs to be changed so that the choice between gain and genetic variability of the next generation become the main focus, not the cut off point approach in selection.

We have provided several examples and compared our method by simulations to the selection methodologies. We have found the optimal genetic mating approach very promising for improving short and long term gains. We believe that successful application of GM will increase the rates of gains per cycle. 

Under the optimal mating breeding scheme some concepts in statistical genetics like selection intensity will have to be adopted so that the choice between gain and genetic variability of the next generation become the main focus, not the cut off point approach in selection.

It is possible to adjust the GM methodology to work with either phenotypic records or the BV's when there are no marker effect estimates. Where PS is relatively more efficient than GS, mating using BV's and the marker data of the parents might be beneficial. In this manuscript, we have only considered additive effects. It would be desirable to extend the objective function to include effects and variances related to dominance, heterosis and epistasis. The optimization procedures described in this paper can be used to optimize over a variety of objective functions with  hard and soft constraints.

\subsubsection*{Supplementary}
\quad File S1 includes the code used for the simulating the data and applying the breeding schemes. Genetic data is simulated using the CRAN package ''hypred'' \citep{hypred}.  Mixed model software is  publicly available via CRAN (Package EMMREML) \citep{EMMREML}.  The rest of the software were written using C++ and R. 

\bibliographystyle{abbrvnat}

\bibliography{MatingBibliography2}

\begin{thebibliography}{43}
\providecommand{\natexlab}[1]{#1}
\providecommand{\url}[1]{\texttt{#1}}
\expandafter\ifx\csname urlstyle\endcsname\relax
  \providecommand{\doi}[1]{doi: #1}\else
  \providecommand{\doi}{doi: \begingroup \urlstyle{rm}\Url}\fi

\bibitem[Akdemir and Godfrey(2015)]{EMMREML}
Deniz Akdemir and Okeke~Uche Godfrey.
\newblock \emph{EMMREML: Fitting Mixed Models with Known Covariance
  Structures}, 2015.
\newblock URL \url{https://CRAN.R-project.org/package=EMMREML}.
\newblock R package version 3.1.

\bibitem[Allaire(1980)]{allaire1980mate}
FR~Allaire.
\newblock Mate selection by selection index theory.
\newblock \emph{Theoretical and Applied Genetics}, 57\penalty0 (6):\penalty0
  267--272, 1980.

\bibitem[Berg et~al.(2006)Berg, Nielsen, S{\o}rensen, et~al.]{berg2006eva}
Peer Berg, J~Nielsen, Morten~Kargo S{\o}rensen, et~al.
\newblock Eva: Realized and predicted optimal genetic contributions.
\newblock In \emph{Proceedings of the 8th World Congress on Genetics Applied to
  Livestock Production, Belo Horizonte, Minas Gerais, Brazil, 13-18 August,
  2006.}, pages 27--09. Instituto Proci{\^e}ncia, 2006.

\bibitem[Bernardo and Charcosset(2006)]{bernardo2006usefulness}
Rex Bernardo and Alain Charcosset.
\newblock Usefulness of gene information in marker-assisted recurrent
  selection: a simulation appraisal.
\newblock \emph{Crop Science}, 46\penalty0 (2):\penalty0 614--621, 2006.

\bibitem[Brisbane and Gibson(1995)]{brisbane1995balancing}
JR~Brisbane and JP~Gibson.
\newblock Balancing selection response and rate of inbreeding by including
  genetic relationships in selection decisions.
\newblock \emph{Theoretical and Applied Genetics}, 91\penalty0 (3):\penalty0
  421--431, 1995.

\bibitem[Charcosset and Hospital(1997)]{charcosset1997marker}
Alain Charcosset and F~Hospital.
\newblock Marker-assisted introgression of quantitative trait loci.
\newblock \emph{Genetics}, 147\penalty0 (3):\penalty0 1469--1485, 1997.

\bibitem[Clark et~al.(2013)Clark, Kinghorn, Hickey, and van~der
  Werf]{clark2013effect}
Samuel~A Clark, Brian~P Kinghorn, John~M Hickey, and Julius~HJ van~der Werf.
\newblock The effect of genomic information on optimal contribution selection
  in livestock breeding programs.
\newblock \emph{Genetics Selection Evolution}, 45\penalty0 (1):\penalty0 1,
  2013.

\bibitem[Crossa et~al.(2006)Crossa, Burgue{\~n}o, Cornelius, McLaren,
  Trethowan, and Krishnamachari]{crossa2006modeling}
Jose Crossa, Juan Burgue{\~n}o, Paul~L Cornelius, Graham McLaren, Richard
  Trethowan, and Anitha Krishnamachari.
\newblock Modeling genotype$\times$ environment interaction using additive
  genetic covariances of relatives for predicting breeding values of wheat
  genotypes.
\newblock \emph{Crop science}, 46\penalty0 (4):\penalty0 1722--1733, 2006.

\bibitem[Daetwyler et~al.(2007)Daetwyler, Villanueva, Bijma, and
  Woolliams]{daetwyler2007inbreeding}
Hans~D Daetwyler, Beatriz Villanueva, Piter Bijma, and John~A Woolliams.
\newblock Inbreeding in genome-wide selection.
\newblock \emph{Journal of Animal Breeding and Genetics}, 124\penalty0
  (6):\penalty0 369--376, 2007.

\bibitem[Davis et~al.(1991)]{davis1991handbook}
Lawrence Davis et~al.
\newblock \emph{Handbook of genetic algorithms}, volume 115.
\newblock Van Nostrand Reinhold New York, 1991.

\bibitem[Falconer et~al.(1996)Falconer, Mackay, and
  Frankham]{falconer1996introduction}
Douglas~S Falconer, Trudy~FC Mackay, and Richard Frankham.
\newblock Introduction to quantitative genetics (4th edn).
\newblock \emph{Trends in Genetics}, 12\penalty0 (7):\penalty0 280, 1996.

\bibitem[Fern{\'a}ndez et~al.(2001)Fern{\'a}ndez, Toro, and
  Caballero]{fernandez2001practical}
J~Fern{\'a}ndez, MA~Toro, and A~Caballero.
\newblock Practical implementation of optimal management strategies in
  conservation programmes: a mate selection method.
\newblock \emph{Animal biodiversity and conservation}, 24\penalty0
  (2):\penalty0 17--24, 2001.

\bibitem[Gianola and Fernando(1986)]{gianola1986bayesian}
Daniel Gianola and Rohan~L Fernando.
\newblock Bayesian methods in animal breeding theory.
\newblock \emph{Journal of Animal Science}, 63\penalty0 (1):\penalty0 217--244,
  1986.

\bibitem[Goddard(2009)]{goddard2009genomic}
Mike Goddard.
\newblock Genomic selection: prediction of accuracy and maximisation of long
  term response.
\newblock \emph{Genetics}, 136\penalty0 (2):\penalty0 245--257, 2009.

\bibitem[Goldberg(2006)]{goldberg2006genetic}
David~E Goldberg.
\newblock \emph{Genetic algorithms}.
\newblock Pearson Education India, 2006.

\bibitem[Henderson(1984)]{henderson1984applications}
CR~Henderson.
\newblock Applications of linear models in animal breeding (university of
  guelph, guelph, on, canada).
\newblock 1984.

\bibitem[Holland(1973)]{holland1973genetic}
John~H Holland.
\newblock Genetic algorithms and the optimal allocation of trials.
\newblock \emph{SIAM Journal on Computing}, 2\penalty0 (2):\penalty0 88--105,
  1973.

\bibitem[Jannink(2010)]{jannink2010dynamics}
Jean-Luc Jannink.
\newblock Dynamics of long-term genomic selection.
\newblock \emph{Genetics Selection Evolution}, 42\penalty0 (1):\penalty0 35,
  2010.

\bibitem[Jannink et~al.(2010)Jannink, Lorenz, and Iwata]{jannink2010genomic}
Jean-Luc Jannink, Aaron~J Lorenz, and Hiroyoshi Iwata.
\newblock Genomic selection in plant breeding: from theory to practice.
\newblock \emph{Briefings in functional genomics}, page elq001, 2010.

\bibitem[Jansen and Wilton(1985)]{jansen1985selecting}
GB~Jansen and JW~Wilton.
\newblock Selecting mating pairs with linear programming techniques.
\newblock \emph{Journal of dairy science}, 68\penalty0 (5):\penalty0
  1302--1305, 1985.

\bibitem[Kinghorn and Shepherd(1999)]{kinghorn1999mate}
BP~Kinghorn and RK~Shepherd.
\newblock Mate selection for the tactical implementation of breeding programs.
\newblock \emph{Assoc Advmt AnimBreed Genet}, 13:\penalty0 130--133, 1999.

\bibitem[Kinghorn(1998)]{kinghorn1998mate}
Brian~P Kinghorn.
\newblock Mate selection by groups.
\newblock \emph{Journal of dairy science}, 81:\penalty0 55--63, 1998.

\bibitem[Kinghorn(2011)]{kinghorn2011algorithm}
Brian~P Kinghorn.
\newblock An algorithm for efficient constrained mate selection.
\newblock \emph{Genetics Selection Evolution}, 43\penalty0 (1):\penalty0 1,
  2011.

\bibitem[Lande and Thompson(1990)]{lande1990efficiency}
Russell Lande and Robin Thompson.
\newblock Efficiency of marker-assisted selection in the improvement of
  quantitative traits.
\newblock \emph{Genetics}, 124\penalty0 (3):\penalty0 743--756, 1990.

\bibitem[Lander et~al.(2001)Lander, Linton, Birren, Nusbaum, Zody, Baldwin,
  Devon, Dewar, Doyle, FitzHugh, et~al.]{lander2001initial}
Eric~S Lander, Lauren~M Linton, Bruce Birren, Chad Nusbaum, Michael~C Zody,
  Jennifer Baldwin, Keri Devon, Ken Dewar, Michael Doyle, William FitzHugh,
  et~al.
\newblock Initial sequencing and analysis of the human genome.
\newblock \emph{Nature}, 409\penalty0 (6822):\penalty0 860--921, 2001.

\bibitem[Legarra et~al.(2009)Legarra, Aguilar, and
  Misztal]{legarra2009relationship}
Andres Legarra, I~Aguilar, and I~Misztal.
\newblock A relationship matrix including full pedigree and genomic
  information.
\newblock \emph{Journal of dairy science}, 92\penalty0 (9):\penalty0
  4656--4663, 2009.

\bibitem[Leutenegger et~al.(2003)Leutenegger, Prum, G{\'e}nin, Verny,
  Lemainque, Clerget-Darpoux, and Thompson]{leutenegger2003estimation}
Anne-Louise Leutenegger, Bernard Prum, Emmanuelle G{\'e}nin, Christophe Verny,
  Arnaud Lemainque, Fran{\c{c}}oise Clerget-Darpoux, and Elizabeth~A Thompson.
\newblock Estimation of the inbreeding coefficient through use of genomic data.
\newblock \emph{The American Journal of Human Genetics}, 73\penalty0
  (3):\penalty0 516--523, 2003.

\bibitem[Margulies et~al.(2005)Margulies, Egholm, Altman, Attiya, Bader,
  Bemben, Berka, Braverman, Chen, Chen, et~al.]{margulies2005genome}
Marcel Margulies, Michael Egholm, William~E Altman, Said Attiya, Joel~S Bader,
  Lisa~A Bemben, Jan Berka, Michael~S Braverman, Yi-Ju Chen, Zhoutao Chen,
  et~al.
\newblock Genome sequencing in microfabricated high-density picolitre reactors.
\newblock \emph{Nature}, 437\penalty0 (7057):\penalty0 376--380, 2005.

\bibitem[Markowitz(1952)]{markowitz1952portfolio}
Harry Markowitz.
\newblock Portfolio selection.
\newblock \emph{The journal of finance}, 7\penalty0 (1):\penalty0 77--91, 1952.

\bibitem[Metzker(2010)]{metzker2010sequencing}
Michael~L Metzker.
\newblock Sequencing technologies-the next generation.
\newblock \emph{Nature reviews genetics}, 11\penalty0 (1):\penalty0 31--46,
  2010.

\bibitem[Meuwissen et~al.(2001)Meuwissen, Hayes, and Goddard]{Meuwissen1819}
T.~H.~E. Meuwissen, B.~J. Hayes, and M.~E. Goddard.
\newblock Prediction of total genetic value using genome-wide dense marker
  maps.
\newblock \emph{Genetics}, 157\penalty0 (4):\penalty0 1819--1829, 2001.

\bibitem[Meuwissen(1997)]{meuwissen1997maximizing}
TH~Meuwissen.
\newblock Maximizing the response of selection with a predefined rate of
  inbreeding.
\newblock \emph{Journal of animal science}, 75\penalty0 (4):\penalty0 934--940,
  1997.

\bibitem[Piepho et~al.(2008)Piepho, M{\"o}hring, Melchinger, and
  B{\"u}chse]{piepho2008blup}
HP~Piepho, J~M{\"o}hring, AE~Melchinger, and A~B{\"u}chse.
\newblock Blup for phenotypic selection in plant breeding and variety testing.
\newblock \emph{Euphytica}, 161\penalty0 (1-2):\penalty0 209--228, 2008.

\bibitem[Pryce et~al.(2012)Pryce, Hayes, and Goddard]{pryce2012novel}
JE~Pryce, BJ~Hayes, and ME~Goddard.
\newblock Novel strategies to minimize progeny inbreeding while maximizing
  genetic gain using genomic information.
\newblock \emph{Journal of dairy science}, 95\penalty0 (1):\penalty0 377--388,
  2012.

\bibitem[Quaas(1988)]{quaas1988additive}
RL~Quaas.
\newblock Additive genetic model with groups and relationships.
\newblock \emph{Journal of Dairy Science}, 71\penalty0 (5):\penalty0
  1338--1345, 1988.

\bibitem[Schierenbeck et~al.(2011)Schierenbeck, Pimentel, Tietze, K{\"o}rte,
  Reents, Reinhardt, Simianer, and K{\"o}nig]{schierenbeck2011controlling}
S~Schierenbeck, ECG Pimentel, M~Tietze, J~K{\"o}rte, R~Reents, F~Reinhardt,
  H~Simianer, and S~K{\"o}nig.
\newblock Controlling inbreeding and maximizing genetic gain using
  semi-definite programming with pedigree-based and genomic relationships.
\newblock \emph{Journal of dairy science}, 94\penalty0 (12):\penalty0
  6143--6152, 2011.

\bibitem[Sonesson et~al.(2012)Sonesson, Woolliams, and
  Meuwissen]{sonesson2012genomic}
Anna~K Sonesson, John~A Woolliams, and Theo~HE Meuwissen.
\newblock Genomic selection requires genomic control of inbreeding.
\newblock \emph{Genetics Selection Evolution}, 44\penalty0 (1):\penalty0 1,
  2012.

\bibitem[Sun et~al.(2013)Sun, VanRaden, O'Connell, Weigel, and
  Gianola]{sun2013mating}
Chuanyu Sun, PM~VanRaden, JR~O'Connell, KA~Weigel, and Daniel Gianola.
\newblock Mating programs including genomic relationships and dominance
  effects.
\newblock \emph{Journal of dairy science}, 96\penalty0 (12):\penalty0
  8014--8023, 2013.

\bibitem[Technow(2014)]{hypred}
Frank Technow.
\newblock \emph{hypred: Simulation of Genomic Data in Applied Genetics}, 2014.
\newblock R package version 0.5.

\bibitem[VanRaden(2008)]{vanraden2008efficient}
PM~VanRaden.
\newblock Efficient methods to compute genomic predictions.
\newblock \emph{Journal of dairy science}, 91\penalty0 (11):\penalty0
  4414--4423, 2008.

\bibitem[Wang(2011)]{wang2011coancestry}
Jinliang Wang.
\newblock Coancestry: a program for simulating, estimating and analysing
  relatedness and inbreeding coefficients.
\newblock \emph{Molecular ecology resources}, 11\penalty0 (1):\penalty0
  141--145, 2011.

\bibitem[Wray and Goddard(1994)]{wray1994moet}
NR~Wray and ME~Goddard.
\newblock Moet breeding schemes for wool sheep 1. design alternatives.
\newblock \emph{Animal Production}, 59\penalty0 (01):\penalty0 71--86, 1994.

\bibitem[Zhong and Jannink(2007)]{zhong2007using}
Shengqiang Zhong and Jean-Luc Jannink.
\newblock Using quantitative trait loci results to discriminate among crosses
  on the basis of their progeny mean and variance.
\newblock \emph{Genetics}, 177\penalty0 (1):\penalty0 567--576, 2007.

\end{thebibliography}

\end{document}